\documentclass[8.5pt,twoside,twocolumn]{article}
\usepackage{graphicx} 
\usepackage{siunitx} 
\usepackage[T1]{fontenc}
\usepackage{pslatex}

\begin{document}

\twocolumn[
  \begin{@twocolumnfalse}
\noindent\LARGE{\textbf{Coherent manipulation of spin qubits based on polyoxometalates: the case of
the single ion magnet [GdW$_{30}$P$_5$O$_{110}$]$^{12-}$ 
$^\dag$}}
\vspace{0.6cm}

\noindent\large{\textbf{Jos\'e J. Baldov\'\i,\textit{$^{a}$}
Salvador Cardona-Serra,\textit{$^{a}$}
Juan M. Clemente-Juan,\textit{$^{a}$}
Eugenio Coronado,$^{\ast}$\textit{$^{a}$}
\\Alejandro Gaita-Ari\~no,\textit{$^{a}$}
Helena Prima-Garc\'\i a$^{\ast}$\textit{$^{a}$}
}}
\vspace{0.5cm}

 \end{@twocolumnfalse} \vspace{0.6cm}
  ]

\noindent\textbf{Polyoxometalate single ion magnet
[GdW$_{30}$P$_5$O$_{110}$]$^{12-}$ (1) has been studied by generalized Rabi
oscillation experiments. It was possible to increase the number of coherent
rotations tenfold through matching the Rabi frequency with the frequency of the
proton. Achieving high coherence with polyoxometalate chemistry, we show their
excellent potential not only for the storage of quantum information but
even for the realization of quantum algorithms.}

\section*{}
\vspace{-1cm}

\footnotetext{\textit{$^{a}$~Instituto de Ciencia Molecular, University of
Valencia, c/Cat. Jos\'e Beltr\'an, 46980, Paterna, Spain.}}

The ultimate state of the art of the miniaturization limit of nanospintronics
is the manipulation of a single electron spin.~\cite{wolf01} 
Single molecule magnets are perfect examples of such control and constitute the
building blocks of molecular spintronics and quantum computing from the
chemistry point of view.~\cite{refMS,refQC} 
Since molecular spins are quantum objects and not just classical
binary memories, the greatest challenge is precisely the manipulation of this
single spin during a sufficiently long time.
In the terms of quantum computing, this means the preservation of quantum
coherence, i.e. all the information of the wave function, during the application of
many quantum gate operations. This is a daunting task, but fortunately
chemistry can provide
the basics for the rational design and optimization of the building-blocks with
the aimed quantum behavior.~\cite{RatDesign}

Among these building-blocks, magnetic polyoxometalates (POMs) combine a rich
magnetochemistry with an arbitrarily low abundance of nuclear spins. 
From this point of view, these molecular metal oxides open the possibility of
developing a large variety of molecular spintronic devices with a hardware
intrinsically suited to preserve the electron spins quantum
coherence.~\cite{CSR2012} Indeed, the preservation of spin coherence is a
formidable challenge for quantum computing applications.~\cite{refQC}$^{(b)}$
Moreover, decoherence is a problem of fundamental importance in physics, with
practical impact of the relaxation processes in chemistry and
engineering.~\cite{decoherence}

In this context, pulsed EPR is an excellent tool for the study of the coherent
manipulation of magnetic systems.~\cite{Ardavan} Under ideal conditions,
which include the absence of microwave radiation or under a short pulse sequence,
the theory estimates that relaxation times of magnetic molecules are in the
order of $\tau_2=$100-500 \si{\micro\second}.~\cite{V15Fe8} 
In real conditions these times are drastically reduced, specially when the
spins are manipulated. The simplest quantum manipulation is known as Rabi
oscillation and consists of a full two-way cycle between states A and B induced
by a microwave field at the transition energy $E_{AB}$. The experimental
observation of these Rabi oscillations is therefore indicative of long
coherence times. In fact, the number of Rabi oscillations is directly connected
to the number of quantum operations that can be performed on the system, so
strategies to extend this number are necessary.

\begin{figure}[htbp]
\includegraphics[width=0.5\columnwidth]{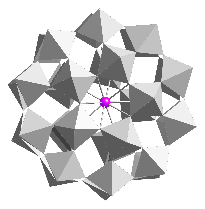}
\raisebox{0.4cm}{\includegraphics[width=0.3\columnwidth]{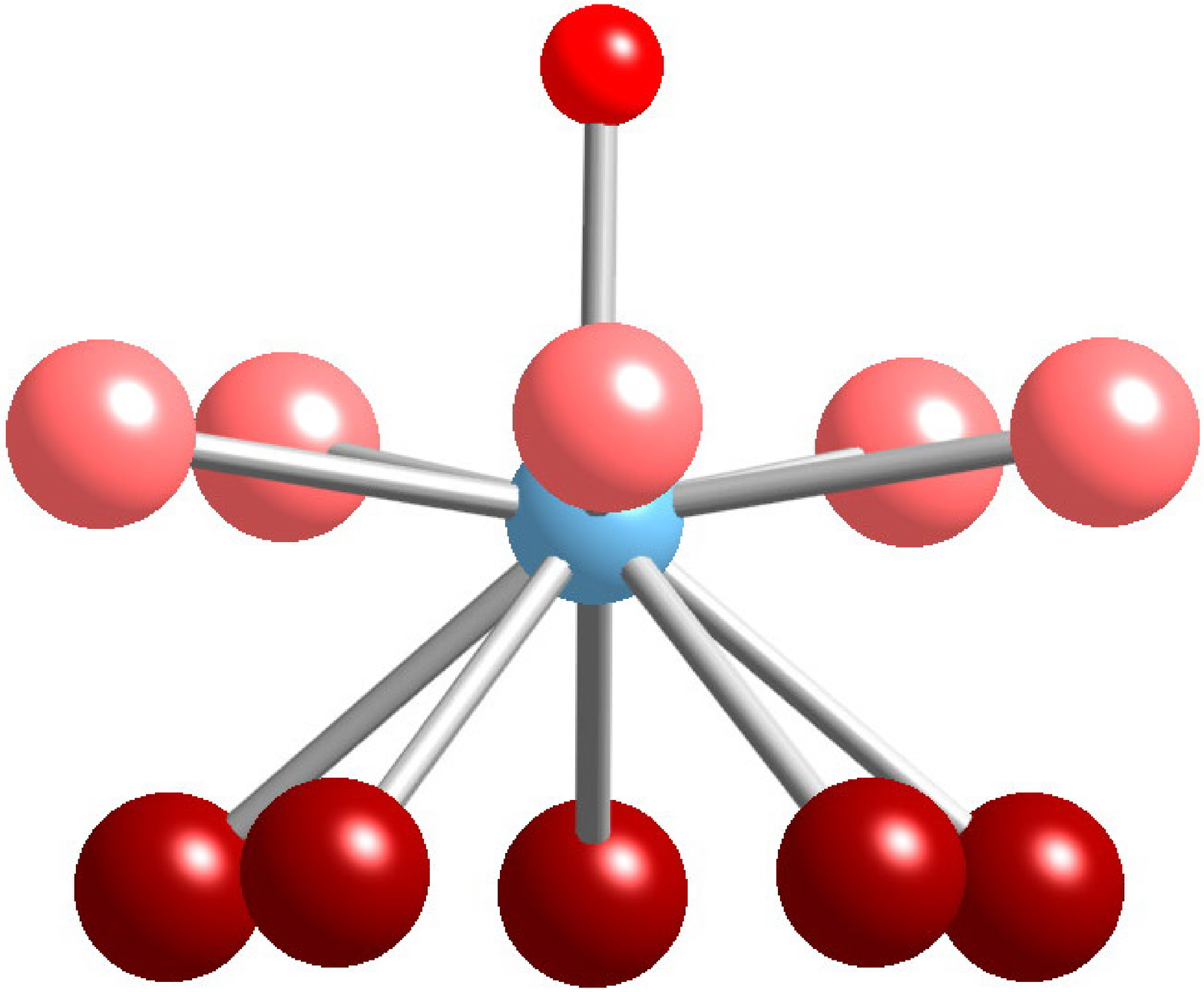} }
\caption{View of {\bf 1} from above (left) and the first coordination sphere of
Gd$^{3+}$ from the side (right). The axial H$_2$O molecule is at 2.2
\AA, which is at least 2.5 times closer than crystallization water molecules}
\label{structure}
\end{figure}

In a recent study of the molecular magnet
[V$_{15}^{IV}$As$_6^{III}$O$_{42}$(H$_2$O]$^{6-}$ (V$_{15}$), the decay time
$\tau_R$ was found to depend strongly on the microwave
power.~\cite{ShimPRL2012} Indeed, two effects were characterized: (i) a linear
dependence of the Rabi decay rate $1/\tau_R$ and the Rabi frequency $\Omega_R$
(or, equivalently, the magnetic component of the microwave field) and (ii) a
large increase in the decay rate near or below the Larmor frequency of the
proton. Effect (i) is a well-known mechanism associated with the dispersion of
the Land\' e $g$ factors of the magnetic molecules and with intercluster
dipolar interactions, but (ii) constituted a new phenomenon related with
dissipative electron-nuclear cross-relaxation. In this communication we extend
this kind of study to [GdW$_{30}$P$_5$O$_{110}$]$^{12-}$ (Fig.~\ref{structure}),
a single ion magnet recently reported by us.~\cite{PRLGdW30} Our
results allow to confirm these two effects and to find a new effect,
characterized by a dramatic enhancement in the coherence.

The studies were made on polycristalline powder of
[YW$_{30}$P$_5$O$_{110}$H$_2$O]K$_{12}\cdot n$H$_2$O doped with 1\%
({\bf 1a}) and 0.1\% ({\bf 1b}) of Gd$^{3+}$.  This is a standard procedure that
weakens Gd-Gd interactions, resulting in longer decoherence times and an easier
observation of Rabi oscillations.  The samples present a broad maximum around
350 mT in echo-induced EPR (Fig. SI1$\dag$), so we chose this region for the
experiments. We varied the microwave power and evaluate the Rabi
frequency $\Omega_R$ and its decay $\tau_R$ (see Mathematical Details in SI$\dag$,
Eq. 1).
This allowed us to qualitatively reproduce, for {\bf 1a} and
{\bf 1b}, the results of reference 9, 
i.e. effects (i) and (ii) in the evolution of the decay rate and Rabi frequency
with different applied microwave powers (see Fig. SI2$\dag$).

\begin{figure}[htbp]
\begin{tabular}{cc}
\includegraphics[width=0.47\columnwidth]{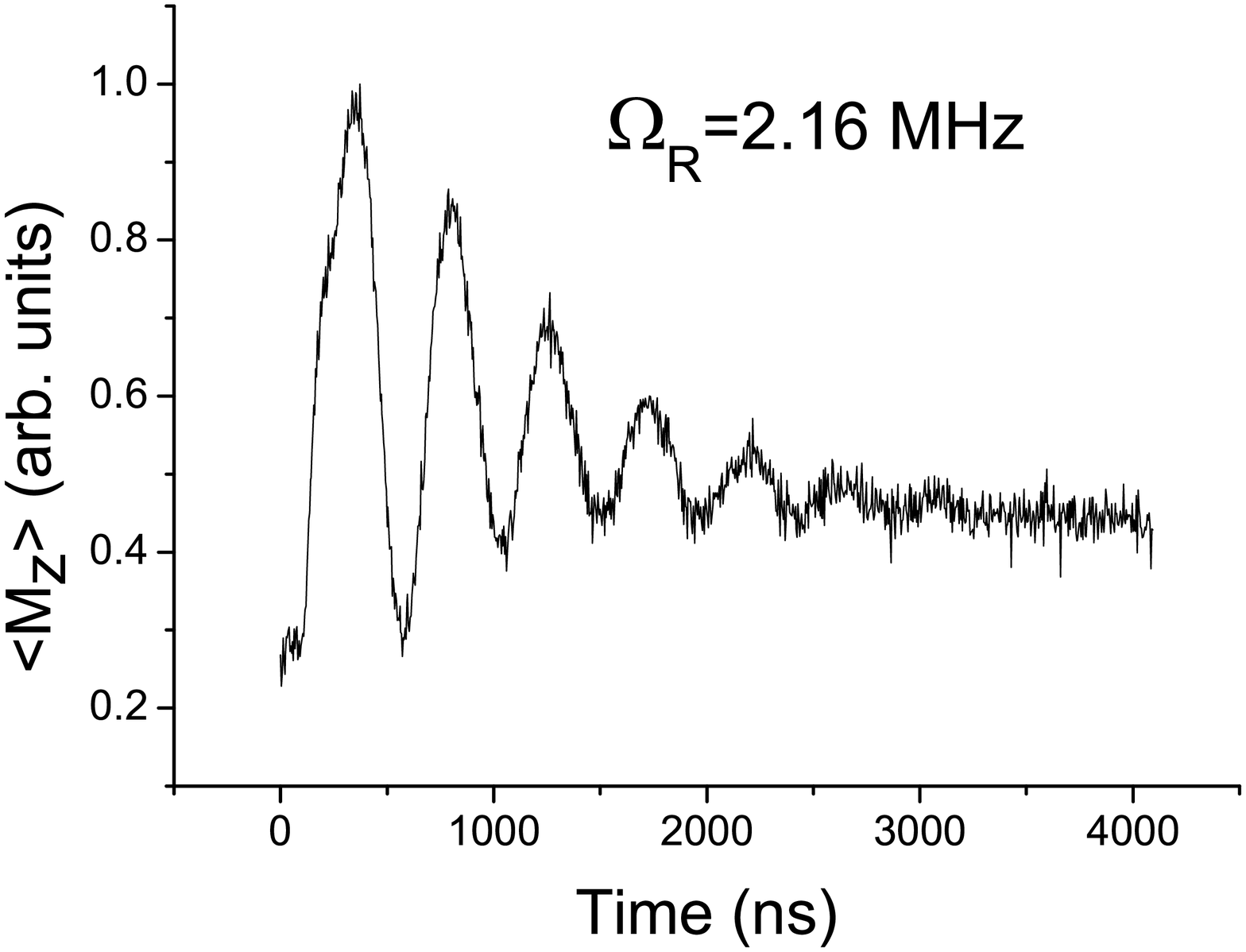}&
\includegraphics[width=0.47\columnwidth]{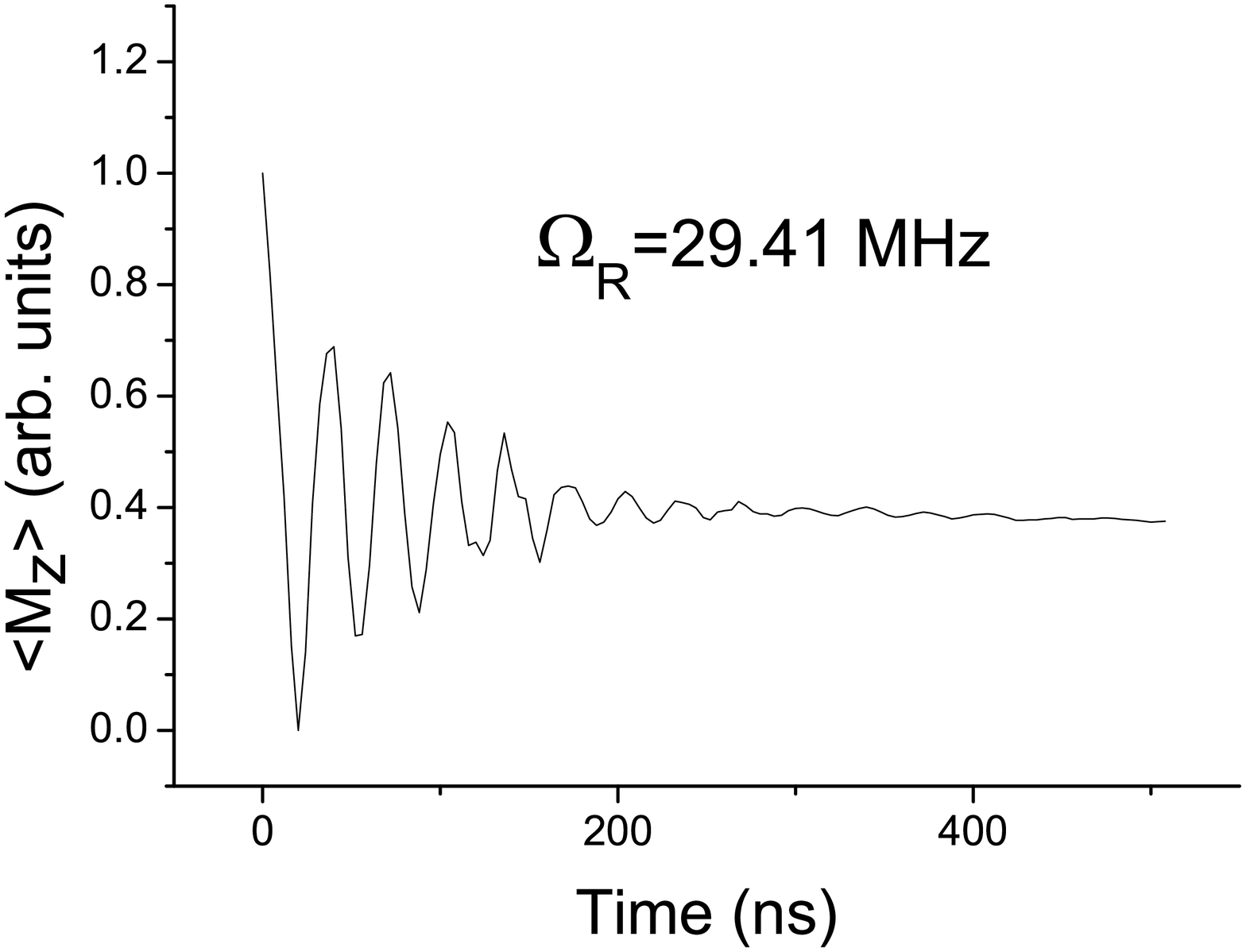}\\
(a)&(b)\\
\end{tabular}
\newline
\includegraphics[width=0.98\columnwidth]{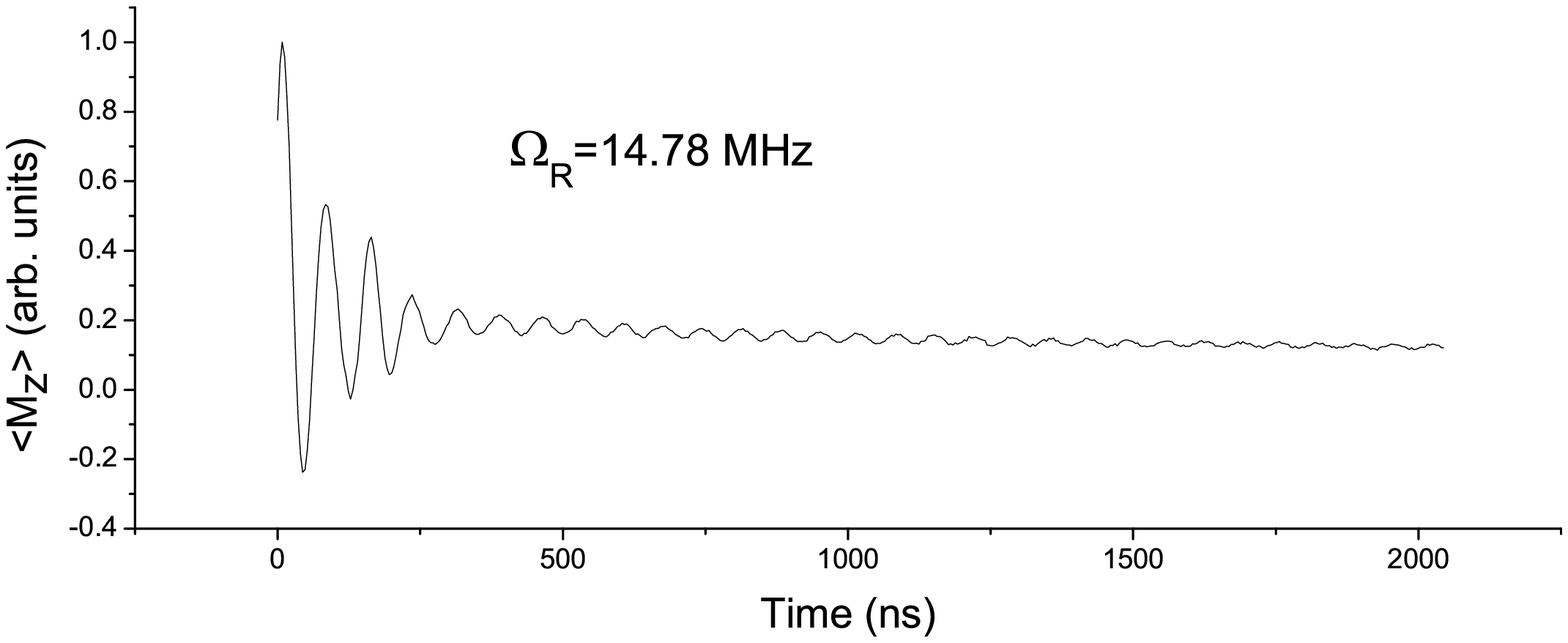}
\centerline{(c)}
\caption{Measured $\langle M_z(t)\rangle$ showing the result of nutation
experiments on {\bf 1a} at microwave powers that are (a) below, (b) above and
(c) near values where the Rabi frequency coincides with the Larmor frequency of
the proton}
\label{resumen}
\end{figure}

Beyond these two effects, an abnormally high number ($>80$) of low-amplitude
oscillations near 15 MHz are observed after the usual decay. This happens 
at long times, i.e. after 500 ns, for both compounds (Fig.~\ref{resumen}(c) and
SI3$\dag$). This effect is clearly observed by comparison of panels (a) and (b)
in Fig.~\ref{resumen}. One can see that while the time scale changes by
approximately an order of magnitude, the number of oscillations remains almost
constant. In contrast, the coherence time on 2(c) is close to that on 2(a) but
with a Rabi frequency in the order of 2(b). Note that at the working field
$B_0=349.6$ mT, the Larmor frequency of the proton is $\nu_H=14.89$ MHz, and
the hyperfine coupling of the Gd$^{3+}$ ground state $^8S_{7/2}$ is also in
this range.~\cite{hfc}

To elucidate the reason behind the extended coherence in a particular frequency
range, we measured coherent oscillations (a) at different applied fields $B_0$
and (b) at different microwave attenuations ($L_{dB}$), which in turn result in
different microwave fields $B_1$. The hyperfine coupling of Gd$^{+3}$ would be
field-independent, so if this was relevant to the phenomenon the frequency
would be expected to be constant and independent of both $B_0$ and $B_1$. 

\begin{figure}[htbp]
\begin{center}
\includegraphics[width=0.61\columnwidth]{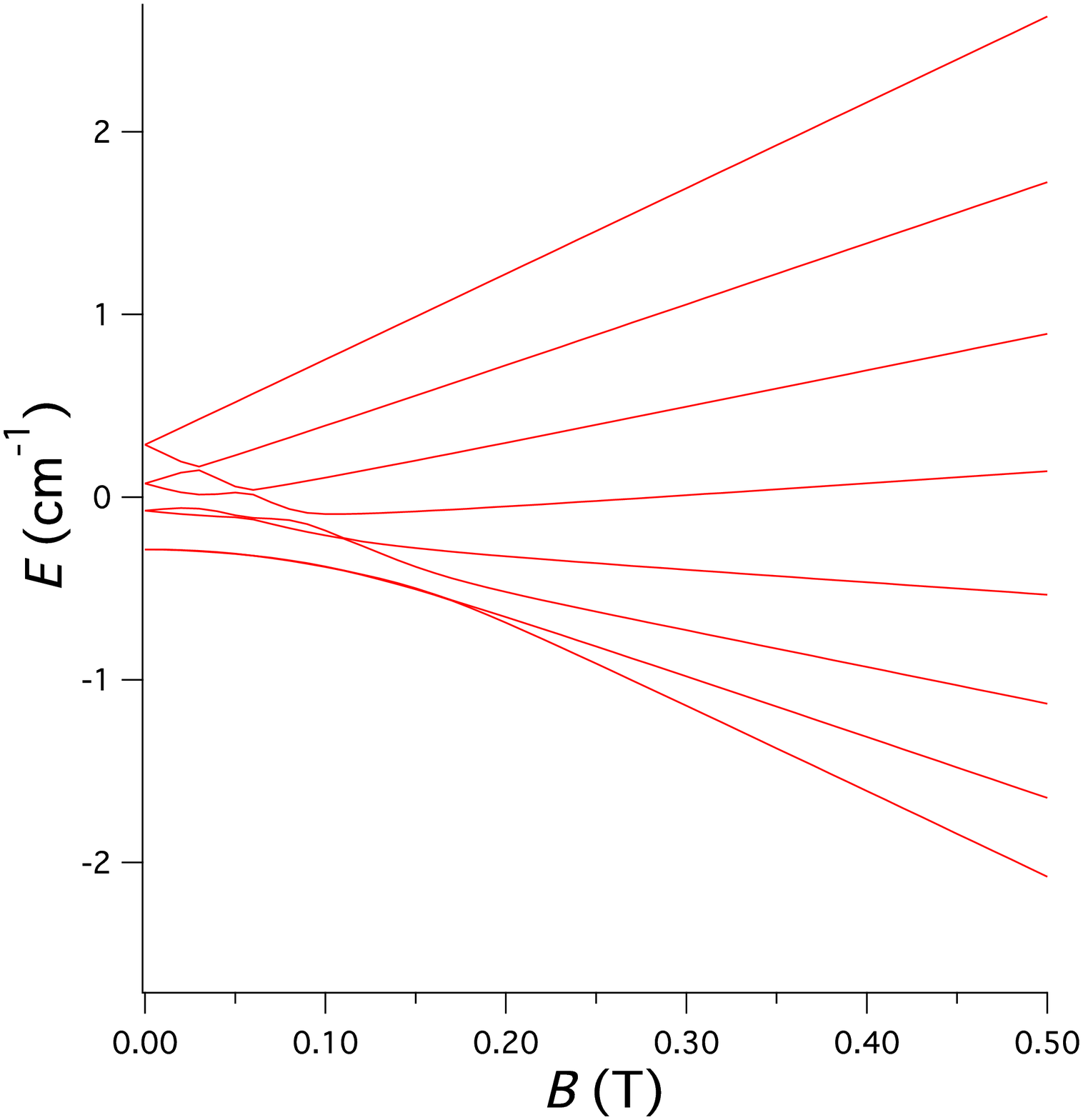}
\hspace{3mm}
\includegraphics[width=0.27\columnwidth]{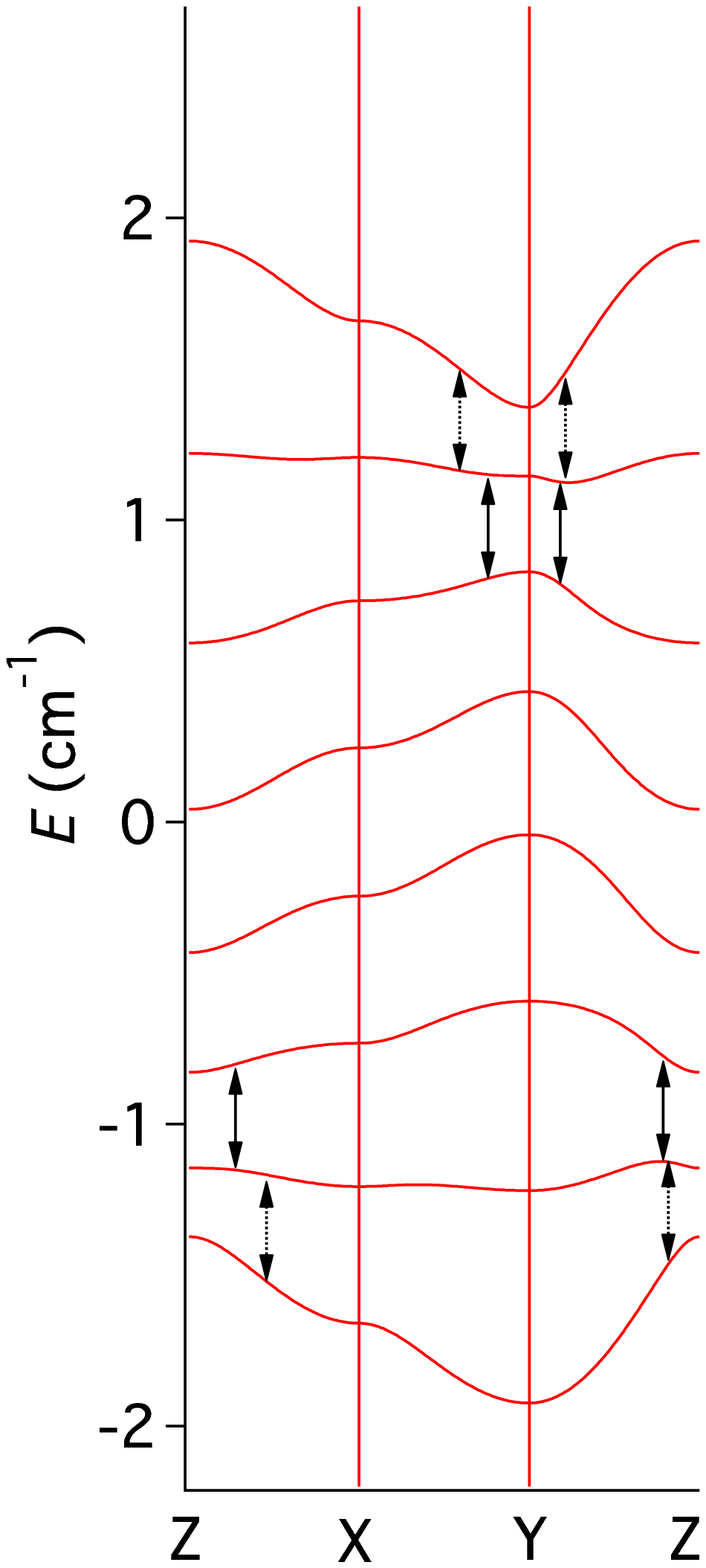}
\end{center}
\caption{Calculated Zeeman splitting of {\bf 1}, sweeping the magnitude of
$B_z$ parallel to the molecular symmetry axis (left) and the orientation of
$|B|=349.6$ mT (right)}
\label{zeeman}
\end{figure}

For a better understanding of the variation of the observed frequency, it is
useful to assign the electronic transitions. The energy level
scheme, as a function of the magnitude and orientation of $B_0$,
(Fig.~\ref{zeeman}) was calculated using crystal field parameters determined in
a previous work,~\cite{PRLGdW30} and introduced in the SIMPRE code.~\cite{SIMPRE}  At 
$B_0=349.6$ mT and for a microcrystalline powder sample the main calculated
contributions are transitions with a character that is mainly either
$\pm7/2\leftrightarrow\pm5/2$ or $\pm5/2\leftrightarrow\pm3/2$. 

Because of the dependence of the Rabi frequency on $M_s$ (see Mathematical
Details in SI$\dag$, Eq. 2),
the transitions that take place between $\pm5/2$ and $\pm3/2$ would be
ideally expected to bear Rabi transitions 31\% larger in frequency than the ones
that occur between $\pm7/2$ and $\pm5/2$. In practice, due to the different
composition of their wave functions, it is more likely that four different
frequencies appear within this window. According to the same type of
calculations and due to the different transitions involved, a slightly larger
frequency range would be expected for 300 mT and a reduced range would be
expected for 400 mT, but the average Rabi frequency should not change.

The Rabi frequencies should be proportional to $B_1$. In contrast, the observed
frequency of long-term oscillations is directly proportional to $B_0$, like the
Larmor frequency of the proton (Table~\ref{rabilarmor}). So apparently a match
between the Rabi oscillations and the frequency of the proton, i.e. the
Hartmann-Hahn condition,~\cite{HH} is necessary for observing this phenomenon.

Note that the oscillations reported in Table~\ref{B0B1} are sustained over a
remarkable attenuation range, corresponding to a $B_1$ field deviation higher
than 40\%. However, below a certain microwave power (above a certain microwave
attenuation $L_{dB}$), the long-term behavior eventually disappears. After this
point, the observed oscillation behaves again as expected for a regular Rabi
oscillation, with short decay times and a frequency that is directly
proportional to the microwave magnetic field $B_1$, i.e. that are slaved to
$L_{dB}$. Indeed, we find that at short times the Rabi oscillation
frequencies are proportional to $B_1$ for every recorded spectra.

\begin{table}[htbp]
\small
\caption{Rabi frequency in MHz at different microwave attenuations
$L_{dB}$ and external fields $B_0$ on {\bf 1b}}
\label{rabilarmor}
\begin{tabular*}{0.5\textwidth}{@{\extracolsep{\fill}}cccc}
\hline
         & 300 mT & 349.6 mT & 400 mT \\
\hline
10 dB    & 13.89 &  8.97 &   -    \\
 9 dB    &   -    & 14.90 & 12.5   \\ 
 8 dB    & 13.16 & 14.71 & 17.86  \\ 
 7 dB    & 13.16 & 14.71 & 16.67  \\ 
 6 dB    & 13.16 & 14.71 & 16.67  \\ 
 5 dB    & 13.16 & 14.71 & 16.67  \\ 
\hline
\end{tabular*}
\label{B0B1}
\end{table}

To conclude, the polyoxometalate single ion magnet [GdW$_{30}$P$_5$O$_{110}$]$^{12-}$ is
of special interest as a spin qubit candidate to do sophisticated EPR
manipulations in the context of molecular spintronics and/or quantum
algorithms.  We have found that in the optimal working conditions the long-term
oscillation frequency is governed by the static magnetic field $B_0$
instead of the microwave magnetic field $B_1$. This suggests a mechanism of
coherence transfer between the electron and nuclear spin at the Hartmann-Hahn
condition, which results in a high number of coherent rotations.  This
behavior, which, to the best of our knowledge, has not been described for other
single-molecule magnets, might be related to the unusual proximity of the
apical water to the lanthanide in this polyoxometalate (see
Fig.~\ref{structure}).
To actually implement a useful algorithm it is necessary to
implement a certain number of quantum operations within the coherence time. In
this case, we see that it is possible to perform at least 80 such operations,
which is ten times higher than the usual range for molecular spin qubits
reported in the literature.

{\it Acknowledgments} ---
The present work has been funded by the EU (Project ELFOS and ERC Advanced
Grant SPINMOL), the Spanish MINECO (grants MAT2011-22785 and
the CONSOLIDER project on Molecular Nanoscience CSD 2007-00010 and the Generalitat
Valenciana (Prometeo and ISIC Programmes of excellence). J.J.B. and S.C.-S.
thank the Spanish MECD for FPU predoctoral grants. We thank A. Ardavan, D.
Kaminski, F. Luis and B. Tsukerblat for useful discussion.

\end{document}